\documentclass[12pt,a4paper]{article}
\usepackage{amsmath}
\usepackage{amssymb,amsfonts}
\usepackage[square,comma]{natbib}
\usepackage{latexsym}

\begin{document}
\title{\textsf{New anisotropic models from isotropic solutions}}
\author{S. D. Maharaj\thanks{Author
for correspondence; email: \texttt{maharaj@ukzn.ac.za}; fax: +2731
260 2632} \; and M. Chaisi\thanks{Permanent address: Department of
Mathematics \& Computer Science, National University of Lesotho,
Roma 180, Lesotho; eMail:
\texttt{m.chaisi@nul.ls}} \\
Astrophysics and Cosmology Research Unit\\
School of Mathematical Sciences \\ University of KwaZulu-Natal\\
Durban 4041, South Africa}
\date{}
\maketitle
\begin{abstract}
We establish an algorithm that produces a new solution to the
Einstein field equations, with an anisotropic matter distribution,
from a given seed isotropic solution. The new solution is expressed
in terms of integrals of known functions, and the integration can be
completed in principle. The applicability of this technique is
demonstrated by generating anisotropic isothermal spheres and
 anisotropic constant density Schwarzschild spheres. Both of these
solutions are expressed in closed form in terms of elementary
functions, and this facilitates physical analysis.
\end{abstract}

AMS Classification numbers: 83C15; 85A99
\section{Introduction \label{sec:intro}}

Solutions to the Einstein field equations with anisotropic matter,
where the radial component of the pressure is not the same as the
tangential component, have been studied by a variety of
investigators in recent years. Such solutions are important in
physics as they may be used to model anisotropic stars where a
static spherically symmetric interior is matched to the
Schwarzschild exterior model. These include the treatments of Dev
and Gleiser \cite{DevGleiser,DevGleiser2}, Herrera \textit{et al.}
\cite{HerreraMartin,HerreraTroconis}, Ivanov \cite{Ivanov}, Mak and
Harko \cite{MakHarko2002,MakHarko} and Sharma and Mukherjee
\cite{SharmaMukherjee2002}. As pointed out in these analyses
anisotropy cannot be neglected when studying the critical mass and
redshift of highly compact bodies, and may be an intrinsic component
of boson stars and strange stars.

In generating exact models describing anisotropic matter
distributions ad hoc assumptions are normally made about the forms
of the matter variables or gravitational potentials. In an attempt
to systematically generate exact solutions Maharaj and Maartens
\cite{MaharajMaartens} proposed an algorithm in which the field
equations are written as a first order system of differential
equations. The energy density and radial pressure are then chosen on
physical grounds; the remaining variables then follow from the field
equations. The Maharaj and Maartens algorithm has been used by
Gokhroo and Mehra \cite{gokhroo}, Chaisi and Maharaj
\cite{ChaisiMaharajA,ChaisiMaharajB} amongst others to produce
physically reasonable anisotropic stars. However these models have
the feature that the anisotropy factor is nonzero in the interior of
the star and an isotropic solution cannot be regained. The objective
of this paper is to provide a new algorithm that may be used to
systematically produce anisotropic solutions to the Einstein field
equations from a given isotropic solution. This solution generating
algorithm has the desirable property of yielding a new exact
anisotropic model which includes an isotropic solution, unlike the
earlier algorithm of Maharaj and Maartens. The new solutions
obtained are members of a one parameter family containing the
isotropic case. We illustrate the utility of the algorithm with two
examples that are physically relevant, the isothermal sphere and the
constant density Schwarzschild sphere.

The principal objective of this paper is to show that we can
generate anisotropic relativistic stars from a specified seed
isotropic star. In Section \ref{sec:fieldeqns}, we describe the
basic features of the model for a relativistic anisotropic star, and
present the differential equations, as a first order system,
governing the gravitational field. The algorithm that produces a new
anisotropic solution from a seed isotropic solution is fully derived
in Section \ref{sec:algorithm}, and we present the final result in
the form of a theorem. The isotropic isothermal model is used to
generate a new anisotropic solution in Section \ref{sec:iso}, to
illustrate the theorem. The conventional isothermal solution with a
linear equation of state is contained in the new class of models. As
a second example the interior Schwarzschild model is used to
generate a new anisotropic solution in Section \ref{sec:Schwarz}.
The conventional interior Schwarzschild solution with constant
density is contained in the new class of models. The details of the
integration process for the second example is given in the Appendix
for convenience. For both examples we study in brief the profile of
the anisotropy factor. In Section \ref{sec:Disc}, we briefly
indicate that the solutions found can be related to stellar models
used in astronomy.


\section{Field equations \label{sec:fieldeqns}}

We utilise a representation of the field equations in which only
first order derivatives appear. This helps to simplify the
integration process as pointed out by Chaisi and Maharaj
\cite{ChaisiMaharajA} whose notation and conventions we follow. We
take the line element for static spherically symmetric spacetimes to
be
\begin{eqnarray}
\mbox{d}s^2 & = &
-e^{\nu}\mbox{d}t^2+e^{\lambda}\mbox{d}r^2+r^2\left(\mbox{d}\theta^2+
\sin^2\theta\mbox{d}\phi^2\right)
\label{metric2}
\end{eqnarray}
where $\nu(r)$ and $\lambda(r)$ are arbitrary functions. The
energy-momentum tensor for isotropic matter which is not radiating
has the form
\begin{eqnarray}
T^{ab}=(\mu+p) u^au^b+pg^{ab} \label{Tiso}
\end{eqnarray}
where the energy density $\mu$, and the isotropic pressure $p$ are
measured relative to the four-velocity $u^a=e^{-\nu/2}\delta^a_0 $
which is comoving. We define the mass function as
\begin{eqnarray}
m(r) & = & \frac{1}{2}\int^r_0x^2\mu(x)\mbox{d}x \label{massFun}
\end{eqnarray}
so that $M=m(R)$ is the total mass of a sphere of radius $R$. The
Einstein field equations, with the help of
(\ref{metric2})-(\ref{massFun}) are equivalent to the system
\begin{subequations} \label{EFEs}
\begin{eqnarray}
e^{-\lambda} & = & 1-\frac{2m}{r} \label{EFEs:1} \\
r(r-2m)\nu^\prime & = & p r^3+2m \label{EFEs:2} \\
\left(\mu+p\right)\nu^\prime+2p^\prime & = & 0 \label{EFEs:3}
\end{eqnarray}
\end{subequations}
for isotropic matter distributions. We are using units where the
speed of light ($c$)  and the coupling constant ($8\pi G/c^4$) are
unity.

The energy-momentum tensor for anisotropic matter which is not
radiating has the form
\begin{eqnarray}
T^{ab}=(\mu+p) u^au^b+pg^{ab}+\pi^{ab} \label{Taniso}
\end{eqnarray}
The quantity
$\pi^{ab}=\sqrt{3}S(r)\left(c^ac^b-\frac{1}{3}h^{ab}\right) $ is the
anisotropic stress tensor; the spacelike vector $c^a =
e^{-\lambda/2}\delta^a_1$ is orthogonal to the fluid four-velocity
$u^a=e^{-\nu/2}\delta^a_0 $ and $|S(r)|$ is the magnitude of the
stress tensor. The Einstein field equations, with (\ref{metric2})
and (\ref{Taniso}), can be written as
\begin{subequations} \label{EFEs2}
\begin{eqnarray}
e^{-\lambda} & = & 1-\frac{2m}{r} \label{EFEs2:1} \\
r(r-2m)\nu^\prime & = & p_r r^3+2m \label{EFEs2:2} \\
\left(\mu+p_r\right)\nu^\prime+2p^\prime_r & = &
-\frac{4}{r}\left(p_r-p_\perp\right) \label{EFEs2:3}
\end{eqnarray}
\end{subequations}
for anisotropic matter distributions. The quantities $p_r$ and
$p_\perp$ are the radial and tangential pressures respectively. The
radial pressure \mbox{$p_r=p+2S/\sqrt{3}$} and the tangential
pressure $p_\perp  = p-S/\sqrt{3}$ are not equal for anisotropic
matter. Note that for isotropic matter $S=0$ and $p_r=p_\perp=p$ and
we regain \eqref{EFEs}. The magnitude $S$ provides a measure of
anisotropy.


\section{The Algorithm \label{sec:algorithm}}

In this section we establish a procedure for generating a new
anisotropic solution of the Einstein field equations from a known
isotropic solution. Consider the Einstein field equations
(\ref{EFEs}) with isotropic matter distribution. Suppose an explicit
solution to (\ref{EFEs}) is known where
\begin{equation}
\left(\nu,\lambda,m,p\right)  =  \left(\nu_0,\lambda_0,
m_0,p_0\right)\label{eq:Soln0}
\end{equation}
and the functions $\nu_0,\;\lambda_0,\;m_0\; \mbox{and}\;p_0$ are
explicitly given. Then the equations in (\ref{EFEs}) are satisfied
and we can write
\begin{subequations} \label{eq:fsoln}
\begin{eqnarray}
 e^{-\lambda_0} & = & 1-\frac{2m_0}{r}    \label{eq:fsoln1}\\
  r(r-2m_0)\nu^\prime_0 & = & p_0r^3+2m_0   \label{eq:fsoln2}\\
 \left(\frac{2m_0^\prime}{r^2}+p_0\right)\nu^\prime_0 +
2p_0^\prime & = & 0 \label{eq:fsoln3}
\end{eqnarray}
\end{subequations}
The equations in \eqref{eq:fsoln} correspond to an isotropic
relativistic sphere. Now consider the Einstein field equations
(\ref{EFEs2}) with anisotropic matter distribution. We seek an
explicit solution to the system \eqref{EFEs2}. To this end we
propose the possible solution
\begin{eqnarray}
 \left(\nu,\;\lambda,\;m,\;p_r,\;p_\perp\right) & =
 & \left(\nu_0+\beta(r),\;\lambda_0,\;m_0,\;p_0+\alpha(r),
 \;p_0-\frac{1}{2}\alpha(r)\right)\label{eq:SolProp}
\end{eqnarray}
where $(\nu_0,\lambda_0,m_0,p_0)$ are given by \eqref{eq:Soln0},
$\alpha$ and $\beta$ are arbitrary functions, and we have set
$\alpha=2S/\sqrt{3}$ for convenience. The equations
(\ref{eq:SolProp}) apply to the anisotropic relativistic sphere.

With the specified form (\ref{eq:SolProp}), the system (\ref{EFEs2})
becomes
\begin{subequations} \label{eq:Subst}
\begin{eqnarray}
e^{-\lambda_0} & = & 1-\frac{2m_0}{r}\\
r(r-2m_0)\nu_0^\prime+r(r-2m_0)\beta^\prime  & = & p_0 r^3+\alpha
r^3 + 2m_0 \\
\left(\frac{2m_0^\prime}{r^2}+p_0+\alpha\right)\left(\nu_0^\prime+\beta^\prime\right)
+ 2p_0^\prime+2\alpha^\prime & = & -\frac{6}{r}\alpha
\end{eqnarray}
\end{subequations}
The systems (\ref{eq:fsoln}) and (\ref{eq:Subst}) imply that
\begin{subequations}
\begin{eqnarray} \label{eq:betalpha}
(r-2m_0)\beta^\prime & = & \alpha r^2\label{eq:betalpha:1}\\
\left(\frac{2m_0^\prime}{r^2}+p_0\right)\beta^\prime +
\alpha\left(\nu_0^\prime+\beta^\prime\right) + 2\alpha^\prime & = &
-\frac{6}{r}\alpha \label{eq:betalpha:2}
\end{eqnarray}
\end{subequations}
which is a coupled system of first order equations. It remains to
integrate \eqref{eq:betalpha} and obtain $\alpha$ and $\beta$.

We can write (\ref{eq:betalpha:2}) as
\begin{eqnarray*}
 \left(\frac{2m_0^\prime}{r^2}+p_0\right)\frac{r^2}{r-2m_0} +
\nu_0^\prime+\beta^\prime + 2\frac{\alpha^\prime}{\alpha} & = &
-\frac{6}{r}
\end{eqnarray*}
with the help of \eqref{eq:betalpha:1}. This differential equation
can be integrated to give
\begin{eqnarray}
\alpha & = &
\frac{k}{r^3}\exp\left\{-\frac{1}{2}\left(I_\alpha+\nu_0+\beta\right)\right\}
\label{eq:alpha}
\end{eqnarray}
where $k$ is a constant, and we have set
\begin{eqnarray}
I_\alpha & = &
\int\left(\frac{2m_0^\prime}{r^2}+p_0\right)\frac{r^2}{r-2m_0}\mbox{d}r\label{eq:Ialpha}
\end{eqnarray}
From (\ref{eq:betalpha:1}) and (\ref{eq:alpha}) we generate the
nonlinear differential equation in $\beta$:
\begin{eqnarray*}
e^{\beta/2}\beta^\prime & = & \frac{k}{r(r-2m_0)}
\exp\left\{-\frac{1}{2}\left(I_\alpha+\nu_0\right)\right\}
\end{eqnarray*}
This nonlinear first order differential equation is integrable and
we generate the result
\begin{eqnarray}
\beta & = &
2\ln\left\{\frac{k}{2}I_\beta+\ell\right\}\label{eq:beta}
\end{eqnarray}
where $\ell$ is a constant of integration, and we have set
\begin{eqnarray}
I_\beta & = & \int \frac{\exp
\left\{-\frac{1}{2}\left(I_\alpha+\nu_0\right)\right\}}{r\left(r
-2m_0\right)}\mbox{d}r\label{eq:Ibeta}
\end{eqnarray}
Thus the anisotropic field equations (\ref{EFEs2}) have been
integrated to generate a new exact solution. Note that with
$k=0(=\alpha)$ this algorithm regains the isotropic solution
\eqref{eq:Soln0}. When $k\ne 0$ then the solution is necessarily
anisotropic. The integrations in $I_\alpha$ and $I_\beta$ in
(\ref{eq:Ialpha}) and (\ref{eq:Ibeta}) respectively can be performed
explicitly as $\nu_0$, $p_0$ and $m_0$ are specified in the
isotropic solution functions in (\ref{eq:Soln0}). We make two
observations relating to the constants of integration. These  could
have been introduced at a later stage as the solution is expressed
in terms of indefinite integrals. If $k=0$ then the parameter $\ell$
has to be greater than zero and can be removed by rescaling the time
coordinate.

We summarise our result in terms of the following theorem:

\textsc{Theorem $\mathcal{A}$}: If $(\nu_0,\lambda_0,m_0,p_0)$ is a
given isotropic solution of the Einstein field equations then
$(\nu,\lambda,m,p_r,p_\perp)$=
$(\nu_0+\beta(r),\lambda_0,m_0,p_0+\alpha(r),p_0-\frac{1}{2}\alpha(r))$
is a new anisotropic solution where
\begin{eqnarray*}
\alpha & = &
\frac{k}{r^3}\exp\left\{-\frac{1}{2}\left(\int\left[\frac{2m_0^\prime}{r^2}
+p_0\right]\frac{r^2}{r-2m_0}\mbox{d}r+\nu_0+\beta\right)\right\}\\
\beta & = & 2\ln\left\{\frac{k}{2}\int \frac{\exp
\left\{-\frac{1}{2}\left(\int\left[\frac{2m_0^\prime}{r^2}
+p_0\right]\frac{r^2}{r-2m_0}\mbox{d}r+\nu_0\right)\right\}}{r\left(r-
2m_0\right)}\mbox{d}r+\ell\right\}
\end{eqnarray*}
and $k$ and $\ell$ are constants.


\section{Anisotropic isothermal spheres \label{sec:iso}}

As a first example we illustrate the applicability of Theorem
$\mathcal{A}$ by generating anisotropic isothermal spheres. The line
element for the isothermal model \cite{SaslawMaharaj} is
\begin{eqnarray}
\mbox{d}s^2 & = &
-r^{\frac{4c}{1+c}}\mbox{d}t^2+\left(1+\frac{4c}{\left(1+c\right)^2}\right)\mbox{d}r^2
+r^2\left(\mbox{d}\theta^2+\sin^2\theta\mbox{d}\phi^2\right)\label{eq:isothermalLE}
\end{eqnarray}
where $c$ is a constant. The relevant isotropic functions for
\eqref{eq:isothermalLE} are
\begin{eqnarray}
\left(\nu_0,\lambda_0,m_0,p_0\right) & = & \left(\frac{4c}{1+c}\ln
r,\;\ln\left\{1+\frac{4c}{(1+c)^2}\right\},\right. \nonumber \\&
&\left.\frac{2cr}{4c+(1+c)^2},\;\frac{1}{r^2}\frac{4c^2}{4c+(1+c)^2}\right)
\label{eq:isothermalISOsol}
\end{eqnarray}

The energy density function, that generates the mass function $m_0$,
has the form
\begin{eqnarray}
\mu_0 & = & \frac{4c}{4c+(1+c)^2}\frac{1}{r^2}\label{eq:mu0iso}
\end{eqnarray}
From \eqref{eq:mu0iso} and  $p_0$ in \eqref{eq:isothermalISOsol} we
observe that
\begin{eqnarray}
p_0 & = & c\mu_0 \label{eq:p0mu0}
\end{eqnarray}
which is a linear barotropic equation of state. Isothermal spheres
with $\mu\propto r^{-2}$ and the equation of state \eqref{eq:p0mu0}
arise in both Newtonian and relativistic stars
\cite{ChaisiMaharajA,ChaisiMaharajB}. They have a long history in
astrophysics as an equilibrium approximation to more complicated
systems which are close to a dynamically relaxed state
\cite{saslaw}.

With the isotropic functions (\ref{eq:isothermalISOsol}) it is
possible to evaluate the integrals $I_\alpha$ and $I_\beta$ in
(\ref{eq:Ialpha}) and (\ref{eq:Ibeta}). Then we generate the
expressions
\begin{eqnarray*}
\alpha & = & k\left(\ell -\frac{k}{2}
\frac{4c+(1+c)^2}{(1+c)(1+5c)}r^{-\frac{1+5c}{1+c}}
\right)^{-1}r^{-\frac{3+7c}{1+c}}\\
\beta & = & 2\ln \left\{\ell -\frac{k}{2}
\frac{4c+(1+c)^2}{(1+c)(1+5c)}r^{-\frac{1+5c}{1+c}} \right\}
\end{eqnarray*}
Hence the new line element has the form
\begin{eqnarray}
\mbox{d}s^2 & = & -r^{\frac{4c}{1+c}}\left(\ell -\frac{k}{2}
\frac{4c+(1+c)^2}{(1+c)(1+5c)}r^{-\frac{1+5c}{1+c}}
\right)^2\mbox{d}t^2+\left(1+\frac{4c}{\left(1+c\right)^2}\right)\mbox{d}r^2\nonumber\\
& &
+r^2\left(\mbox{d}\theta^2+\sin^2\theta\mbox{d}\phi^2\right)\label{eq:isothermalLE1}
\end{eqnarray}
and the matter variables have the explicit form
\begin{subequations} \label{eq:isothermalANISOsol}
\begin{eqnarray}
m & = & \frac{2cr}{4c+(1+c)^2}  \\
p_r & = & \frac{1}{r^2}\frac{4c^2}{4c+(1+c)^2} \nonumber\\& &+
k\left(\ell -\frac{k}{2}
\frac{4c+(1+c)^2}{(1+c)(1+5c)}r^{-\frac{1+5c}{1+c}}
\right)^{-1}r^{-\frac{3+7c}{1+c}} \\
p_\perp & = & \frac{1}{r^2}\frac{4c^2}{4c+(1+c)^2}\nonumber\\& &
-\frac{k}{2}\left(\ell -\frac{k}{2}
\frac{4c+(1+c)^2}{(1+c)(1+5c)}r^{-\frac{1+5c}{1+c}}
\right)^{-1}r^{-\frac{3+7c}{1+c}}
\end{eqnarray}
\end{subequations}
The isotropic isothermal sphere model \eqref{eq:isothermalLE}
generates the anisotropic isothermal sphere model
\eqref{eq:isothermalLE1}. With the parameter values $k  = 0,\;\ell =
1$, we regain the conventional isothermal sphere.

\begin{figure}[thb]
\vspace{1.8in} \includegraphics{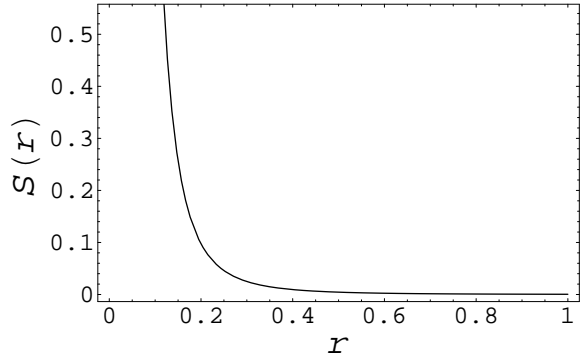} \caption{\label{fig:SisoA}Decreasing
$S(r)$ for anisotropic isothermal sphere; $c= 0.1$, $k= 0.005$, and
$\ell=10$}
\end{figure}

The degree of anisotropy is
\begin{eqnarray}
S & = & \frac{k}{2}\sqrt{3}\left(\ell -\frac{k}{2}
\frac{4c+(1+c)^2}{(1+c)(1+5c)}r^{-\frac{1+5c}{1+c}}
\right)^{-1}r^{-\frac{3+7c}{1+c}}\label{eq:isothermalSA}
\end{eqnarray}
A graph of the anisotropy factor \eqref{eq:isothermalSA} was plotted
with the help of Mathematica \cite{wolfram}. This is given in Figure
\ref{fig:SisoA} for the particular values of the parameters shown.
The anisotropy factor $S$ is plotted against the radial distance on
the interval $0< r\leq 1$. It is worth noting from this graph that
the anisotropy has the feature that it is a monotonically decreasing
function as $r$ approaches the boundary subject to a particular
choice of parameters. There is a singularity at $r=0$ which $S$
shares with the other dynamical and metric functions. However there
are other choices of parameters that could be made such that $S$ is
a monotonically increasing function if the physics of the problem
demanded such behavior.  An increasing profile for $S$ would  be
physically  relevant for boson stars as pointed out by Dev and
Gleiser \cite{DevGleiser}; however note that their analysis was
performed for a constant energy density. The simple behavior of $S$
reflected in this  graph indicates that a full physical
investigation of this solution is possible; we will perform this
investigation in future work.

\section{Anisotropic Schwarzschild spheres\label{sec:Schwarz}}

As a second example we illustrate the utility of Theorem
$\mathcal{A}$ by generating anisotropic Schwarzschild spheres. The
line element for the interior Schwarzschild model \cite{DelgatyLake}
is
\begin{eqnarray}
\mbox{d}s^2 & = &
-\left(A-B\sqrt{1-\frac{r^2}{R^2}}\right)^2\mbox{d}t^2+\left(1-
\frac{r^2}{R^2}\right)^{-1}\mbox{d}r^2\nonumber\\&
&
+r^2\left(\mbox{d}\theta^2+\sin^2\theta\mbox{d}\phi^2\right)\label{eq:schwarzsLE}
\end{eqnarray}
where $A$ and $B$ are constants. The relevant isotropic functions
for \eqref{eq:schwarzsLE} are
\begin{eqnarray}
\left(\nu_0,\lambda_0,m_0,p_0 \right) & = &
\left(2\ln\left\{A-B\sqrt{1-\frac{r^2}{R^2}}\right\},\;-\ln\left\{1-
\frac{r^2}{R^2}\right\},\right.\nonumber\\
& &
\left.\frac{r^3}{2R^2},\;-\frac{1}{R^2}\left(\frac{A-3B\sqrt{1-
\frac{r^2}{R^2}}}{A-B\sqrt{1-\frac{r^2}{R^2}}}\right)\right)
\label{eq:schwarzsISOsols}
\end{eqnarray}

The energy density function, that generates the mass function $m_0$,
has the form $\mu_0=3/R^2$. We therefore have
\begin{eqnarray}
\mu_0 & = & \mbox{constant}\label{eq:schwarzsMU0}
\end{eqnarray}
which essentially replaces the equation of state \eqref{eq:p0mu0} of
Section \ref{sec:iso}. The interior of dense neutron stars and
superdense relativistic stars are of near uniform density
\cite{MaharajLeach,RhoadesRuffini}. Consequently the assumption
\eqref{eq:schwarzsMU0} of uniform energy density is often made in
the modelling process \cite{DevGleiser,MaharajMaartens,BowersLiang}.

The integral $I_\alpha$ in (\ref{eq:Ialpha}) is easily integrated
and we obtain
\begin{eqnarray}
I_\alpha & = & 2\ln\left\{A-B\sqrt{1-\frac{r^2}{R^2}}\right\}
-\ln\left\{1-\frac{r^2}{R^2}\right\}\label{eq:IalphaSchw}
\end{eqnarray}
However the integral $I_\beta$  takes the form
\begin{eqnarray}
I_\beta & = & \int
\frac{1}{r^2\sqrt{1-\frac{r^2}{R^2}}\left(A-B\sqrt{1-\frac{r^2}{R^2}}\right)^2}\mbox{d}r
\label{eq:SchwarzIalphaA}
\end{eqnarray}
The integrand in (\ref{eq:SchwarzIalphaA}) is complicated; however
the integration can be completed and $I_\beta$ can be written purely
in terms of elementary functions. The details of the integration are
given in the Appendix.

We need to investigate two cases: $A\ne B$ and $A=B$. We observe
from the Appendix that $I_\alpha$ and $I_\beta$ can be written in
closed form for these two cases. Two line elements arise and we
represent our solutions as follows:

\emph{Case I}: $A\ne B$

In this case the line element is
\begin{eqnarray}
\mbox{d}s^2 & = &
-\left(A-B\sqrt{1-\frac{r^2}{R^2}}\right)^2\nonumber\\
& & \times\left[\frac{k}{2R\left(A^2-B^2\right)^2}\left(
\frac{6AB^2}{\sqrt{B^2-A^2}}\tanh^{-1}\left\{\frac{(A+B)\tan
\left\{\frac{1}{2}\sin^{-1}\frac{r}{R}\right\}}{\sqrt{B^2
-A^2}}\right\}\right.\right.\nonumber\\
& &
-\frac{R}{r}\left(2AB+\left(A^2+B^2\right)\sqrt{1-\frac{r^2}{R^2}}\right)\nonumber\\
& & \left.\left.
-\frac{B^3r}{R}\left(A-B\sqrt{1-\frac{r^2}{R^2}}\right)^{-1}\right)+\ell\right]^2
\mbox{d}t^2\nonumber\\
& & +\left(1-\frac{r^2}{R^2}\right)^{-1}\mbox{d}r^2
+r^2\left(\mbox{d}\theta^2+\sin^2\theta\mbox{d}\phi^2\right)\label{eq:schwarzsLE1}
\end{eqnarray}
and the matter variables become
\begin{subequations} \label{eq:schwarzsANISOsols}
\begin{eqnarray}
m & = & \frac{r^3}{2R^2}  \\
p_r & = &
-\frac{1}{R^2}\left(\frac{A-3B\sqrt{1-\frac{r^2}{R^2}}}{A-B\sqrt{1
-\frac{r^2}{R^2}}}\right)
+\frac{k\sqrt{1-\frac{r^2}{R^2}}}{r^3\left(A-B\sqrt{1-\frac{r^2}{R^2}}\right)^2}\nonumber\\
& & \times\left[\frac{k}{2R\left(A^2-B^2\right)^2}\left(
\frac{6AB^2}{\sqrt{B^2-A^2}}\tanh^{-1}\left\{\frac{(A
+B)\tan\left\{\frac{1}{2}\sin^{-1}\frac{r}{R}\right\}}{\sqrt{B^2
-A^2}}\right\}\right.\right.\nonumber\\
& &
-\frac{R}{r}\left(2AB+\left(A^2+B^2\right)\sqrt{1-\frac{r^2}{R^2}}\right)\nonumber\\
& & \left.\left.
-\frac{B^3r}{R}\left(A-B\sqrt{1-\frac{r^2}{R^2}}\right)^{-1}\right)+\ell\right]^{-1}
\end{eqnarray}
\begin{eqnarray}
p_\perp & = &
-\frac{1}{R^2}\left(\frac{A-3B\sqrt{1-\frac{r^2}{R^2}}}{A-B\sqrt{1
-\frac{r^2}{R^2}}}\right)
-\frac{k\sqrt{1-\frac{r^2}{R^2}}}{2r^3\left(A-B\sqrt{1
-\frac{r^2}{R^2}}\right)^2}\nonumber\\
& & \times\left[\frac{k}{2R\left(A^2-B^2\right)^2}\left(
\frac{6AB^2}{\sqrt{B^2-A^2}}\tanh^{-1}\left\{\frac{(A
+B)\tan\left\{\frac{1}{2}\sin^{-1}\frac{r}{R}\right\}}{\sqrt{B^2
-A^2}}\right\}\right.\right.\nonumber\\
& &
-\frac{R}{r}\left(2AB+\left(A^2+B^2\right)\sqrt{1-\frac{r^2}{R^2}}\right)\nonumber\\
& & \left.\left.
-\frac{B^3r}{R}\left(A-B\sqrt{1-\frac{r^2}{R^2}}\right)^{-1}\right)+\ell\right]^{-1}
\end{eqnarray}
\end{subequations}
With the parameter values $ k=0,\; \ell=1 $ we regain the original
interior Schwarzschild sphere (\ref{eq:schwarzsLE}).

\begin{figure}[thb]
\vspace{1.8in} \includegraphics{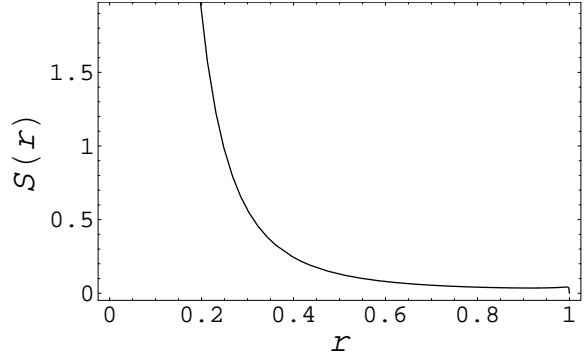}
\caption{\label{fig:SschwarzA}Decreasing $S(r)$ for anisotropic
Schwarzschild sphere ($A\ne B$); $A=1$, \mbox{$B=-10$}, $k=2$,
$\ell=1$ and $R=1$.}
\end{figure}

The degree of anisotropy is
\begin{eqnarray}
S & = & \frac{k\sqrt{3}\sqrt{1-\frac{r^2}{R^2}}}{2r^3\left(A-B\sqrt{1
-\frac{r^2}{R^2}}\right)^2}\nonumber\\
& & \times\left[\frac{k}{2R\left(A^2-B^2\right)^2}\left(
\frac{6AB^2}{\sqrt{B^2-A^2}}\tanh^{-1}\left\{\frac{(A
+B)\tan\left\{\frac{1}{2}\sin^{-1}\frac{r}{R}\right\}}{\sqrt{B^2
-A^2}}\right\}\right.\right.\nonumber\\
& &
-\frac{R}{r}\left(2AB+\left(A^2+B^2\right)\sqrt{1
-\frac{r^2}{R^2}}\right)\nonumber\\
& & \left.\left.
-\frac{B^3r}{R}\left(A-B\sqrt{1-\frac{r^2}{R^2}}\right)^{-1}\right)
+\ell\right]^{-1}\label{eq:schwarzSA}
\end{eqnarray}
The graph of the anisotropy factor \eqref{eq:schwarzSA} was plotted
with the assistance of Mathematica \cite{wolfram}. This is shown in
Figure \ref{fig:SschwarzA} for the particular values of the
parameters. The interval for the plot of $S$ against $r$ is $0<
r\leq 1$. The quantity $S$ is a monotonically decreasing function.
Subject to the choice of the parameters, the anisotropy can be
constructed such that it is a monotonically decreasing function as
$r$ approaches the boundary. The function $S$ vanishes at the
boundary. Other choices of the parameters $A$, $B$, $k$ and $\ell$
may generate different behavior for $S$. The singularity at the
center does not seem to be `removable' by any choice of parameters.
A comparison of this case with Dev and Gleiser \cite{DevGleiser} is
not possible as the anisotropy factor vanishes at the boundary
$r=R$. This will be investigated further and a detailed analysis of
the anisotropy factor $S$ and the dynamical variables for this
anisotropic solution will be pursued in the future.

\newpage

\emph{Case II}: $A=B$

For this case the line element is
\begin{eqnarray}
\mbox{d}s^2 & = & -\left(1-\sqrt{1-\frac{r^2}{R^2}}\right)^2\nonumber\\
& & \times\left[\ell -\frac{k}{80R}\csc^5
\left\{\frac{1}{2}\sin^{-1}\frac{r}{R}\right\}
\sec\left\{\frac{1}{2}\sin^{-1}\frac{r}{R}\right\}\right.\nonumber\\
& &\left.
\times\left(\left(3-\frac{2r^2}{R^2}\right)\sqrt{1
-\frac{r^2}{R^2}}-2\left(1-\frac{2r^2}{R^2}\right)\right)
 \right]^2\mbox{d}t^2\nonumber\\
& &
+\left(1-\frac{r^2}{R^2}\right)^{-1}\mbox{d}r^2
+r^2\left(\mbox{d}\theta^2+\sin^2\theta\mbox{d}\phi^2\right)
\end{eqnarray}
and the matter variables are given by
\begin{subequations} \label{eq:SchwarzA=B}
\begin{eqnarray}
m & = & \frac{r^3}{2R^2}\\
p_r & = &
-\frac{1}{R^2}\left(\frac{1-3\sqrt{1-\frac{r^2}{R^2}}}{1
-\sqrt{1-\frac{r^2}{R^2}}}\right)\nonumber\\
& & +
\frac{k\sqrt{1-\frac{r^2}{R^2}}}{r^3\left(1-\sqrt{1
-\frac{r^2}{R^2}}\right)^2}
\left[\ell -\frac{k}{80R}\csc^5\left\{\frac{1}{2}\sin^{-1}
\frac{r}{R}\right\}\sec\left\{\frac{1}{2}\sin^{-1}\frac{r}{R}\right\}\right.\nonumber\\
& &\left.
\times\left(\left(3-\frac{2r^2}{R^2}\right)\sqrt{1-
\frac{r^2}{R^2}}-2\left(1-\frac{2r^2}{R^2}\right)\right)
\right]^{-1}\\
 p_\perp & = &
-\frac{1}{R^2}\left(\frac{1-3\sqrt{1-\frac{r^2}{R^2}}}{1
-\sqrt{1-\frac{r^2}{R^2}}}\right)\nonumber\\
& & -
\frac{k\sqrt{1-\frac{r^2}{R^2}}}{2r^3\left(1-\sqrt{1-\frac{r^2}{R^2}}\right)^2}
\left[\ell -\frac{k}{80R}\csc^5\left\{\frac{1}{2}
\sin^{-1}\frac{r}{R}\right\}\sec\left\{\frac{1}{2}
\sin^{-1}\frac{r}{R}\right\}\right.\nonumber\\
& &\left.
\times\left(\left(3-\frac{2r^2}{R^2}\right)\sqrt{1
-\frac{r^2}{R^2}}-2\left(1-\frac{2r^2}{R^2}\right)\right)
 \right]^{-1}
\end{eqnarray}
\end{subequations}
With the parameter values $ k  =  0,\; \ell  =  1 $, we regain the
original interior Schwarzschild sphere \eqref{eq:schwarzsLE} with
$A=B$.

\begin{figure}[thb]
\vspace{1.9in} \includegraphics{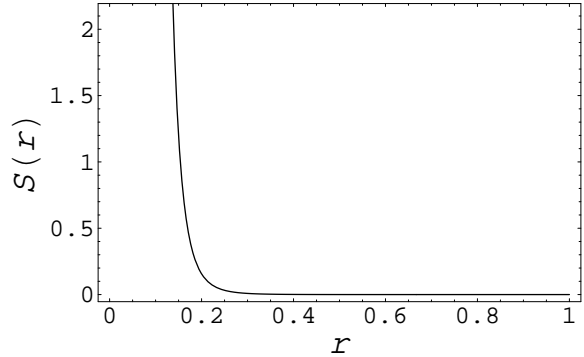}
\caption{\label{fig:SschwarzAB}Decreasing $S(r)$ for anisotropic
Schwarzschild sphere ($A=B$); $k=0.00006$, $\ell=100$ and $R=1$}
\end{figure}

The degree of anisotropy is
\begin{eqnarray}
S & = &
\frac{k\sqrt{3}\sqrt{1-\frac{r^2}{R^2}}}{2r^3\left(1-\sqrt{1-\frac{r^2}{R^2}}\right)^2}
\left[ \ell -\frac{k}{80R}\csc^5\left\{\frac{1}{2}
\sin^{-1}\frac{r}{R}\right\}\sec\left\{\frac{1}{2}\sin^{-1}
\frac{r}{R}\right\}\right.\nonumber\\
& &\left.
\times\left(\left(3-\frac{2r^2}{R^2}\right)\sqrt{1
-\frac{r^2}{R^2}}-2\left(1-\frac{2r^2}{R^2}\right)\right)
\right]^{-1}\label{eq:SchwarzSAB}
\end{eqnarray}
The graph of the anisotropy factor \eqref{eq:SchwarzSAB} was plotted
with the assistance of Mathematica \cite{wolfram}. This is shown in
Figure \ref{fig:SschwarzAB} for the particular values of the
parameters. The interval for the plot of $S$ against $r$ is $0<
r\leq 1$. The quantity $S$ is a monotonically decreasing function.
The behavior of $S$ is similar to the case $A\ne B$ given in Figure
\ref{fig:SschwarzA}. However observe that the behavior in Figure
\ref{fig:SschwarzAB} is more restricted for \emph{Case II} as $A$
and $B$ are fixed.

\section{Discussion\label{sec:Disc}}

It is remarkable that the integrations in Theorem $\mathcal{A}$ can
be completed in closed form in spite of the complicated nature of
the integrands in $\alpha$ and $\beta$, and we obtain expressions in
terms of elementary functions. It is possible to demonstrate that
the solutions found can be used to discuss the structure of neutron
stars and quasi-stellar objects. Now consider a neutron star of
radius 10 kilometres and surface density of $2\times
10^{14}\;\mbox{gcm}^{-3}$ in c.g.s units. Then the parameters
 relating to the surface redshift $z=\left(1-2M/R\right)^{-1/2}-1$, and
 the mass $M$ in
terms of the solar masses $M_\odot$ can be calculated. We choose
values so that comparison with Gokhroo and Mehra \cite{gokhroo} is
facilitated. For the choice $c= 0.1,$ we obtain  $z= 0.154$ and $M=
0.520 M_\odot$.  The choice $c= 0.9,$ generates the values $z=
0.413$ and $M= 1.40 M_\odot$. Clearly other values for $z$ and $M$
can be generated by making different choices for $c$ and the surface
density. The range of values obtainable is consistent with the
results of Gokhroo and Mehra \cite{gokhroo}. Hence our solutions
yield values for surface redshifts and masses that correspond to
realistic stellar sources such as Her X-1 and Vela X-1.

\textbf{\large Acknowledgements}

SDM and MC thank the National Research Foundation of South Africa
for financial support. MC is grateful to the University of
KwaZulu-Natal for a scholarship. We are grateful to the referees for
valuable comments.


\section*{Appendix}

The integral $I_\beta$ has the form
\begin{eqnarray*}
I_\beta & = & \int
\frac{1}{r^2\sqrt{1-\frac{r^2}{R^2}}\left(A-B\sqrt{1-\frac{r^2}{R^2}}\right)^2}\mbox{d}r
\end{eqnarray*}
Note that two cases arise in the integration process: $A\ne B$ and
$A=B$.

\emph{Case I}: $A\ne B$

To carry out the integration in \eqref{eq:SchwarzIalphaA} for $A\ne
B$ we make the substitution
\begin{eqnarray*}
\sin\vartheta=\frac{r}{R}
\end{eqnarray*}
so that $I_\beta$ becomes
\begin{eqnarray*}
I_\beta & = & \frac{1}{R}\int\frac{1}{\sin^2\vartheta(A
-B\cos\vartheta)^2}\mbox{d}\vartheta \\
& = & \frac{1}{R\left(A^2-B^2\right)^2}\left(
\frac{6AB^2}{\sqrt{B^2-A^2}}\tanh^{-1}\left\{\frac{(A+B)
\tan\frac{\vartheta}{2}}{\sqrt{B^2-A^2}}\right\}\right.\\
& &\left.
-\left(2AB+A^2\cos\vartheta+B^2\cos\vartheta\right)\csc\vartheta
-\frac{B^3\sin\vartheta}{A-B\cos\vartheta} \right)
\end{eqnarray*}
in terms of $\vartheta$. In terms of the original radial coordinate
$r$ we have
\begin{eqnarray*}
 I_\beta
& = & \frac{1}{R\left(A^2-B^2\right)^2}\left(
\frac{6AB^2}{\sqrt{B^2-A^2}}\tanh^{-1}\left\{\frac{(A+B)\tan
\left\{\frac{1}{2}\sin^{-1}\frac{r}{R}\right\}}{\sqrt{B^2-A^2}}\right\}\right.\\
& & \left.
-\frac{R}{r}\left(2AB+\left(A^2+B^2\right)\cos\left\{\sin^{-1}\frac{r}{R}\right\}\right)
-\frac{B^3r}{R\left(A-B\cos\left\{\sin^{-1}\frac{r}{R}\right\}\right)}\right)\\
& = & \frac{1}{R\left(A^2-B^2\right)^2}\left(
\frac{6AB^2}{\sqrt{B^2-A^2}}\tanh^{-1}\left\{\frac{(A+B)\tan
\left\{\frac{1}{2}\sin^{-1}\frac{r}{R}\right\}}{\sqrt{B^2-A^2}}\right\}\right.\\
& & \left.
-\frac{R}{r}\left(2AB+\left(A^2+B^2\right)\sqrt{1-\frac{r^2}{R^2}}\right)
-\frac{B^3r}{R}\left(A-B\sqrt{1-\frac{r^2}{R^2}}\right)^{-1}\right)
\end{eqnarray*}
Then \eqref{eq:alpha} and \eqref{eq:beta} give
\begin{eqnarray*}
\alpha & = &
\frac{k\sqrt{1-\frac{r^2}{R^2}}}{r^3\left(A-B\sqrt{1-\frac{r^2}{R^2}}\right)^2}\\
& & \times\left[\frac{k}{2R\left(A^2-B^2\right)^2}\left(
\frac{6AB^2}{\sqrt{B^2-A^2}}\tanh^{-1}\left\{\frac{(A+B)\tan
\left\{\frac{1}{2}\sin^{-1}\frac{r}{R}\right\}}{\sqrt{B^2-A^2}}\right\}\right.\right.\\
& &
-\frac{R}{r}\left(2AB+\left(A^2+B^2\right)\sqrt{1-\frac{r^2}{R^2}}\right)\\
& & \left.\left.
-\frac{B^3r}{R}\left(A-B\sqrt{1-\frac{r^2}{R^2}}\right)^{-1}
\right)+\ell\right]^{-1} \end{eqnarray*}
\begin{eqnarray*} \beta & =
& 2\ln\left\{\frac{k}{2R\left(A^2-B^2\right)^2}\left(
\frac{6AB^2}{\sqrt{B^2-A^2}}\tanh^{-1}\left\{\frac{(A+B)\tan
\left\{\frac{1}{2}\sin^{-1}\frac{r}{R}\right\}}{\sqrt{B^2-A^2}}\right\}\right.\right.\\
& &
-\frac{R}{r}\left(2AB+\left(A^2+B^2\right)\sqrt{1-\frac{r^2}{R^2}}\right)\\
& & \left.\left.
-\frac{B^3r}{R}\left(A-B\sqrt{1-\frac{r^2}{R^2}}\right)^{-1}\right)+\ell\right\}
\end{eqnarray*}
for the functions $\alpha$ and $\beta$.
%
%

\emph{Case II}: $A=B$

To carry out the integration in \eqref{eq:SchwarzIalphaA} with $A=B$
we again make the substitution
\begin{eqnarray*}
\sin\vartheta & = & \frac{r}{R}
\end{eqnarray*}
to have
\begin{eqnarray*}
I_\beta & = &\int\frac{1}{r^2\sqrt{1-\frac{r^2}{R^2}}\left(1-\sqrt{1
-\frac{r^2}{R^2}}\right)^2}\mbox{d}r\\
& = & \frac{1}{R}\int\frac{1}{\sin^2\vartheta\left(1-\cos\vartheta\right)^2}
\mbox{d}\vartheta\\
& = &
-\frac{1}{80R}\csc^5\frac{\vartheta}{2}\sec\frac{\vartheta}{2}
\left(5\cos\vartheta-4\cos2\vartheta+\cos3\vartheta\right)
\end{eqnarray*}
The terms $\cos 2\vartheta$ and $\cos 3\vartheta$ can be simplified
with basic trigonometric identities and $I_\beta$, in terms of the
original radial coordinate $r$, becomes
\begin{eqnarray*}
I_\beta & = & -\frac{1}{80R}\csc^5\left\{\frac{1}{2}\sin^{-1}
\frac{r}{R}\right\}\sec\left\{\frac{1}{2}\sin^{-1}\frac{r}{R}\right\}\\
& & \times\left(5\sqrt{1-\frac{r^2}{R^2}}
-4\left(1-\frac{r^2}{R^2}-\frac{r^2}{R^2}\right)+\sqrt{1
-\frac{r^2}{R^2}}\left(1-\frac{r^2}{R^2}-3\frac{r^2}{R^2}\right)\right) \\
& = & -\frac{1}{80R}\csc^5\left\{\frac{1}{2}\sin^{-1}\frac{r}{R}\right\}
\sec\left\{\frac{1}{2}\sin^{-1}\frac{r}{R}\right\}\\
& &
\times\left(\left(6-\frac{4r^2}{R^2}\right)\sqrt{1-\frac{r^2}{R^2}}-4\left(1
-\frac{2r^2}{R^2}\right)\right)
\end{eqnarray*}
Then \eqref{eq:alpha} and \eqref{eq:beta} give
\begin{eqnarray*}
\alpha & = & \frac{k\sqrt{1-\frac{r^2}{R^2}}}{r^3\left(1-\sqrt{1
-\frac{r^2}{R^2}}\right)^2} \left[\ell -\frac{k}{80R}\csc^5\left\{\frac{1}{2}
\sin^{-1}\frac{r}{R}\right\}\sec\left\{\frac{1}{2}\sin^{-1}\frac{r}{R}\right\}\right.\\
& &\left.
\times\left(\left(3-\frac{2r^2}{R^2}\right)\sqrt{1-\frac{r^2}{R^2}}-2
\left(1-\frac{2r^2}{R^2}\right)\right) \right]^{-1}\\
\beta & = & 2\ln\left\{\ell -\frac{k}{80R}\csc^5\left\{\frac{1}{2}\sin^{-1}
\frac{r}{R}\right\}\sec\left\{\frac{1}{2}\sin^{-1}\frac{r}{R}\right\}\right.\\
& &\left.
\times\left(\left(3-\frac{2r^2}{R^2}\right)\sqrt{1
-\frac{r^2}{R^2}}-2\left(1-\frac{2r^2}{R^2}\right)\right)
\right\}
\end{eqnarray*}
for the functions $\alpha$ and $\beta$.
%
%
{}

%
\end{document}